\documentclass[10pt,conference,letterpaper]{IEEEtran}

\usepackage{amsfonts}
\usepackage{amsmath}
\usepackage{amssymb}
\usepackage{booktabs}
\usepackage{balance}
\usepackage{color}
\usepackage{citesort}
\usepackage{graphicx}
\usepackage{mathrsfs}
\usepackage{mdwmath}
\usepackage{multirow}
\usepackage{setspace}
\usepackage{subfigure}

\ifCLASSINFOpdf
\else
\fi

\newcommand{\CarN}{N}
\newcommand{\CarK}{K}
\newcommand{\CarBlocks}{M}
\newcommand{\basicN}{n}
\newcommand{\basicK}{k}
\newcommand{\Memory}{m}
\newcommand{\diag}{\mbox{diag}}
\newcommand{\Qfun}[1]{{\rm Q}\left(#1\right)}

\ifCLASSOPTIONonecolumn
\newcommand{\FigOneWidth}{0.8\textwidth}
\newcommand{\FigTwoWidth}{0.7\textwidth}
\newcommand{\FigThreeWidth}{0.7\textwidth}
\fi
\ifCLASSOPTIONtwocolumn
\newcommand{\FigOneWidth}{0.45\textwidth}
\newcommand{\FigTwoWidth}{0.43\textwidth}
\newcommand{\FigThreeWidth}{0.44\textwidth}
\fi
\newcommand{\vspaceh}{-0.1cm}

\newcommand{\mathbfit}[1]{\mbox{\boldmath$#1$\unboldmath}}

\newtheorem{algorithm}{\textbf{Algorithm}}
\newtheorem{example}{\textbf{Example}}
\makeatletter
    
    \newcommand{\Rmnum}[1]{\expandafter\@slowromancap\romannumeral #1@}
\makeatother

\ifCLASSOPTIONtwocolumn
\renewcommand{\baselinestretch}{0.96}
\fi

\IEEEoverridecommandlockouts
\begin{document}
\title{EXIT Chart Analysis of Block Markov Superposition Transmission of Short Codes}
%\huge
%
%\author{
%    \IEEEauthorblockN{Kechao Huang\IEEEauthorrefmark{1},
%    Xiao Ma\IEEEauthorrefmark{1}, and~Daniel~J.~Costello,~Jr.\IEEEauthorrefmark{2}}
%
%    \IEEEauthorblockA{\IEEEauthorrefmark{1}Dept. of ECE, Sun Yat-sen University, Guangzhou, China, hkech@mail2.sysu.edu.cn, maxiao@mail.sysu.edu.cn}
%
%    \IEEEauthorblockA{\IEEEauthorrefmark{2}Dept. of EE, University of Notre Dame, Notre Dame, USA, Email: dcostel1@nd.edu}
%    \thanks{This work was partially supported by the $973$ Program (No. $2012$CB$316100$), the China NSF (No. 61172082 and No. 91438101), and the U.S. NSF (No. CCF-1161754).}
% \vspace{-0.1cm}
%}

\ifCLASSOPTIONonecolumn
\author{
    \IEEEauthorblockN{Kechao Huang\IEEEauthorrefmark{1},
    Xiao Ma\IEEEauthorrefmark{1}, and~Daniel~J.~Costello,~Jr.\IEEEauthorrefmark{2}}

    \IEEEauthorblockA{\IEEEauthorrefmark{1}Dept. of ECE, Sun Yat-sen University, Guangzhou, China\\ Email: hkech@mail2.sysu.edu.cn, maxiao@mail.sysu.edu.cn}

    \IEEEauthorblockA{\IEEEauthorrefmark{2}Dept. of EE, University of Notre Dame, Notre Dame, USA\\ Email: dcostel1@nd.edu}
% \vspace{-0.45cm}
}
\fi
\ifCLASSOPTIONtwocolumn
\author{
  \IEEEauthorblockN{Kechao Huang and Xiao Ma}
  \IEEEauthorblockA{Dept. of ECE, Sun Yat-sen University\\
    Guangzhou 510006, GD, China\\
    Email: maxiao@mail.sysu.edu.cn}
  \and
  \IEEEauthorblockN{Daniel~J.~Costello,~Jr.}
  \IEEEauthorblockA{Dept. of EE, University of Notre Dame\\
    Notre Dame 46556, Indiana, USA\\
    Email: dcostel1@nd.edu}
    \thanks{This work was partially supported by the $973$ Program (No. $2012$CB$316100$), the China NSF (No. 61172082 and No. 91438101), and the U.S. NSF (No. CCF-1161754).}
}
%
%\fi

\maketitle

\begin{abstract}
To be considered for a 2015 IEEE Jack Keil Wolf ISIT Student Paper Award. In this paper, a modified extrinsic information transfer~(EXIT) chart analysis that takes into account the relation between mutual information~(MI) and bit-error-rate~(BER) is presented to study the convergence behavior of block Markov superposition transmission~(BMST) of short codes~(referred to as basic codes). We show that the threshold curve of BMST codes using an iterative sliding window decoding algorithm with a fixed decoding delay achieves a lower bound in the high signal-to-noise ratio (SNR) region, while in the low SNR region, due to error propagation, the thresholds of BMST codes become slightly worse as the encoding memory increases. We also demonstrate that the threshold results are consistent with finite-length performance simulations.

\end{abstract}
%
%\begin{IEEEkeywords}
%Iterative message processing/passing algorithm, low-density parity-check~(LDPC) convolutional codes, sliding window decoding, stopping criteria.
%\end{IEEEkeywords}

\section{Introduction}
Spatially coupled low-density parity-check (SC-LDPC) codes are constructed by coupling together a series of $L$ disjoint Tanner graphs of an underlying LDPC block code~(LDPC-BC) into a single coupled chain and can be viewed as a type of LDPC convolutional code~\cite{Felstrom99}. It was shown in~\cite{Lentmaier10,Mitchell14} that the belief propagation~(BP) decoding thresholds of SC-LDPC code ensembles are numerically indistinguishable from the maximum {\em a posteriori}~(MAP) decoding thresholds of their underlying LDPC-BC ensembles. Subsequently, it was proven analytically that SC-LDPC code ensembles exhibit {\em threshold saturation} on memoryless binary-input symmetric-output channels under BP decoding~\cite{Kudekar13}. Due to their excellent performance, SC-LDPC codes have received a great deal of attention recently~(see,~e.g.,~\cite{Pusane11,Iyengar12,Andriyanova13,Mitchell13,Wei14_IT,Huang14} and the references therein).
%It is well known that SC-LDPC code ensembles exhibit a phenomenon called ``threshold saturation"~\cite{Lentmaier10,Kudekar11}, i.e., the belief propagation~(BP) decoding threshold saturates to the maximum a-posteriori (MAP) threshold of the corresponding uncoupled LDPC-BC ensemble.Costello14,

The concept of spatial coupling is not limited to LDPC codes. Block Markov superposition transmission~(BMST) of short codes~\cite{Ma13x,Liang14c}, for example, is equivalent to spatial coupling of the subgraphs that specify the generator matrices of the short codes. From this perspective, BMST codes are similar to braided block codes~\cite{Feltstrom09}, staircase codes~\cite{Smith12}, and spatially coupled turbo codes~\cite{Moloudi14}. A BMST code can also be viewed as a serially concatenated code with a structure similar to repeat-accumulate-like codes~\cite{Abbasfar07}. The outer code is a short code, referred to as the {\em basic code}~(not limited to repetition codes), that introduces redundancy, while the inner code is a rate-one block-oriented feedforward convolutional code~(instead of a bit-oriented accumulator) that introduces memory between transmissions. Hence, BMST codes typically have very simple encoding algorithms. To decode BMST codes, a sliding window decoding algorithm with a tunable decoding delay can be used, as with SC-LDPC codes. The construction of BMST codes is flexible~\cite{Liang14x}, in the sense that it applies to all code rates of interest in the interval (0,1). Further, BMST codes have near-capacity performance~(observed by simulation) in the waterfall region of the bit-error-rate~(BER) cruve and an error floor~(predicted by analysis) that can be controlled by the encoding memory.
%An interesting question is how to predict the waterfall region of the bit-error-rate~(BER) cruve.
%A BMST code can also be treated as a serially concatenated code, whose encoding structure is similar to that of the repeat-accumulate-like codes~\cite{Abbasfar07}. The outer code is a short code~(referred to as the \emph{basic codes} and not limited to the repetition code~(RC)) that introduces redundancy, while the inner code is a block-oriented feedforward rate-one convolutional code~(instead of a bit-oriented accumulator) that introduces memory among transmissions. As with SC-LDPC codes, a sliding window decoding algorithm with a tunable decoding delay can be used to decode BMST codes. Compared with SC-LDPC codes, the construction of BMST codes is flexible, in the sense that they achieve near capacity performance in the waterfall region with easy encoding process, which can be almost as fast as that for the basic code. Further, BMST codes have an error floor which can be controlled by the encoding memory. Since the error floor of BMST codes is predictable, an interesting question is how to predict the waterfall region of the bit-error-rate~(BER) curve.

On an additive white Gaussian noise channel~(AWGNC), the well-known extrinsic information transfer~(EXIT) chart analysis~\cite{Brink01} can be used to obtain the threshold of LDPC-BC ensembles. In~\cite{Liva07}, a novel EXIT chart analysis was used to evaluate the performance of protograph-based LDPC-BC ensembles, and a similar analysis was used to find the thresholds of $q$-ary SC-LDPC codes with sliding window decoding in~\cite{Wei14_IT}. Unlike LDPC codes, the asymptotic  BER of BMST codes with window decoding cannot be better than a corresponding genie-aided lower bound~\cite{Ma13x}. Thus, conventional EXIT chart analysis cannot be applied directly to BMST codes. In this paper, we propose a modified EXIT chart analysis, that takes into account the relation between mutual information~(MI) and BER, to study the convergence behavior of BMST codes and to predict the performance in the waterfall region of the BER curve. We also show that the modified EXIT chart analysis of BMST codes is supported by finite-length performance simulations.

\section{SC-LDPC Codes vs. BMST Codes}\label{SecII}
In this section, both SC-LDPC codes and BMST codes are described in terms of matrices for the purpose of showing their similarities~(dualities) and differences.
\subsection{Protograph-Based SC-LDPC Codes}
A protograph-based SC-LDPC code ensemble can be constructed from a protograph-based LDPC-BC code ensemble using the edge spreading technique~\cite{Mitchell14}, described here in terms of the {\em base~(parity-check) matrix} representation of code ensembles. Let $\mathbfit{B}$ be a $(\CarN-\CarK)\times \CarN$ base matrix representing an LDPC-BC ensemble with design rate $R=\CarK/\CarN$. A terminated SC~(convolutional) base matrix $\mathbfit{B}_{\rm SC}$ with \emph{coupling width~(syndrome former memory)} $\Memory$ and \emph{coupling length} $L$ can be constructed by applying the edge spreading technique to $\mathbfit{B}$, resulting in
%\vspace{\vspaceh}
\begin{eqnarray}\label{SC-LDPC_ter}
\mathbfit{B}_{\rm SC}=
    \left[
    \begin{array}{cccc}
        \mathbfit{B}_0     &                 &       & \\
        \mathbfit{B}_1     &\mathbfit{B}_0     &       & \\
        \vdots           &\mathbfit{B}_1     &\ddots & \\
        \mathbfit{B}_{\Memory} &\vdots           &\ddots &\mathbfit{B}_0\\
                         &\mathbfit{B}_{\Memory} &\ddots &\mathbfit{B}_1\\
                         &                 &\ddots &\vdots\\
                         &                 &       &\mathbfit{B}_{\Memory}\\
    \end{array}
    \right],
\end{eqnarray}
where the $\Memory+1$ component submatrices $\mathbfit{B}_0,\mathbfit{B}_1,\ldots,\mathbfit{B}_{\Memory}$, each of size $(\CarN-\CarK) \times \CarN$, satisfy $\sum\limits_{i=0}^{\Memory}\mathbfit{B}_i =\mathbfit{B}$. The graph lifting operation is then applied to $\mathbfit{B}_{\rm SC}$ by replacing each nonzero entry in $\mathbfit{B}_{\rm SC}$ with a randomly selected $M \times M$ permutation matrix\footnote{If the nonzero entry $B_{i,j}>1$, it is replaced by a summation of $B_{i,j}$ nonoverlapping randomly selected permutation matrices of size $M \times M$.} and each zero entry in $\mathbfit{B}_{\rm SC}$ with the $M \times M$ all-zero matrix,
resulting in a terminated SC-LDPC code with constraint length $v_s = (\Memory+1)M\CarN$, where $M$~(typically a large integer) is the {\em lifting factor}. The resulting SC-LDPC parity-check matrix $\mathbfit{H}_{\rm SC}$ of size $(L+\Memory)(\CarN-\CarK)M \times L\CarN M$ is given by
\vspace{\vspaceh}
\begin{align*}
\mathbfit{H}_{\rm SC}= ~~~~~~~~~~~~~~~~~~~~~~~~~~~~~~~~~~~~~~~~~~~~~~~~~~~~ \nonumber \\
    \begin{bmatrix}
    %\begin{smallmatrix}
        \mathbfit{H}_0(0)     &                    &       & \\
        \mathbfit{H}_1(1)     &\mathbfit{H}_0(1)     &       & \\
        \vdots              &\mathbfit{H}_{1}(2)   &\ddots & \\
        \mathbfit{H}_{\Memory}(\Memory)   &\vdots              &\ddots &\mathbfit{H}_0(L-1)\\
                            &\mathbfit{H}_{\Memory}(\Memory+1) &\ddots &\mathbfit{H}_1(L)\\
                            &                    &\ddots &\vdots\\
                            &                    &       &\mathbfit{H}_{\Memory}(L+\Memory-1)\\
    %\end{smallmatrix}
    \end{bmatrix},
\end{align*}
where the blank spaces in $\mathbfit{H}_{\rm SC}$ correspond to zeros and the submatrices $\mathbfit{H}_i(t)$ have size $(\CarN-\CarK)M \times \CarN M$, $\forall i,t$. The design rate of the terminated SC-LDPC code ensemble is given by
\vspace{\vspaceh}
\begin{eqnarray}\label{R_SC-LDPC}
    R_{\rm SC} = 1-\frac{(L+\Memory)(\CarN-\CarK)}{L\CarN}=1-\frac{L+\Memory}{L}(1-R),
\end{eqnarray}
which is slightly less than the design rate $R=\CarK/\CarN$ of the uncoupled LDPC-BC ensemble due to the termination. However, this rate loss becomes vanishingly small as $L\rightarrow \infty$.

\subsection{BMST Codes}
In contrast to SC-LDPC codes, it is convenient to describe BMST codes using generator matrices. To describe a BMST code ensemble with coupling width~(encoding memory) $\Memory$ and coupling length $L$, we start with an $L \times (L+\Memory)$ matrix
\vspace{\vspaceh}
\begin{align}\label{G_basic}
\mathbfit{A} =
        \begin{bmatrix}
           %第一行
           1 &1 &\cdots &1 & &  \\
             &1 &1      &\cdots  &1 &  \\
             &  &\ddots &\ddots  &\ddots &\ddots &\\
             &  &       &1       &1      &\cdots &1 \\
             &  &       &        &1      &1 &\cdots &1 \\
        \end{bmatrix},
\end{align}
which has constant weight $\Memory+1$ in each row. Now assuming that we want to construct a rate $R=\basicK/\basicN$ code, we select a basic code with a $\basicK \times \basicN$ generator matrix $\mathbfit{G}$. Let $\mathbfit{\varPi}_i$ $(0 \leq i \leq \Memory)$ be $\Memory+1$ randomly selected $\basicN \times \basicN$ permutation matrices. Then each nonzero entry $A_{i,j}$ in $\mathbfit{A}$ is replaced with a $\basicK \times \basicN$ matrix $\mathbfit{G}\mathbfit{\varPi}_{j-i}$ and each zero entry in $\mathbfit{A}$ is replaced with the $\basicK \times \basicN$ all-zero matrix, resulting in a BMST code of length $(L+\Memory)\basicN$ and dimension $L\basicK$. The resulting generator matrix $\mathbfit{G}_{\rm BMST}$ of the BMST code is given by
\vspace{\vspaceh}
\begin{align*}\label{G_BMST}
\mathbfit{G}_{\rm BMST} = ~~~~~~~~~~~~~~~~~~~~~~~~~~~~~~~~~~~~~~~~~~~~~~~~~~~~~~~ \nonumber \\
    \begin{bmatrix}\begin{smallmatrix}
           %第一行
           \mathbfit{G}\mathbfit{\varPi}_{0}       & \mathbfit{G}\mathbfit{\varPi}_{1} & \cdots &
           \mathbfit{G}\mathbfit{\varPi}_{\Memory} &                    &  \\
           %第二行
                              & \mathbfit{G}\mathbfit{\varPi}_{0}       & \mathbfit{G}\mathbfit{\varPi}_{1} &
           \cdots             & \mathbfit{G}\mathbfit{\varPi}_{\Memory} &  \\
           %第三行
                              &                     & \ddots &
           \ddots             & \ddots              & \ddots  &\\
           %第四行
                              &                     &        &
           \mathbfit{G}\mathbfit{\varPi}_{0}       & \mathbfit{G}\mathbfit{\varPi}_{1}  & \cdots &
           \mathbfit{G}\mathbfit{\varPi}_{\Memory} \\
           %第五行
                              &                     &        &
                              & \mathbfit{G}\mathbfit{\varPi}_{0}        & \mathbfit{G}\mathbfit{\varPi}_{1} &
           \cdots             & \mathbfit{G}\mathbfit{\varPi}_{\Memory} \\
        \end{smallmatrix}\end{bmatrix}.
\end{align*}
The rate of the BMST code is
\vspace{\vspaceh}
\begin{eqnarray}\label{R_BMST}
    R_{\rm BMST} = \frac{L \basicK}{(L+\Memory)\basicN} = \frac{L}{L+\Memory}R,
\vspace{\vspaceh}
\end{eqnarray}
which is slightly less than the rate $R=\basicK/\basicN$ of the basic code. However, similar to SC-LDPC codes, this rate loss becomes vanishingly small as $L\rightarrow \infty$.

Though any code~(linear or nonlinear) with a fast encoding algorithm and an efficient soft-in soft-out~(SISO) decoding algorithm can be taken as the basic code, in this paper we focus on the use of the $\CarBlocks$-fold Cartesian product of a repetition code~(RC) or a single parity-check~(SPC) code as the basic code, resulting in a BMST-RC code or a BMST-SPC code, respectively. Let $\mathbfit{G}_0$ be the $\CarK \times \CarN$ generator matrix of an RC code or an SPC code. The $\basicK \times \basicN$ generator matrix $\mathbfit{G}$ of the basic code is then given by
\vspace{\vspaceh}
\begin{equation}\label{G_basic_code}
\mathbfit{G} =
  \diag\{\underbrace{\mathbfit{G}_0, \cdots, \mathbfit{G}_0}_\CarBlocks \},
\vspace{\vspaceh}
\end{equation}
where $\diag\left\{ \mathbfit{G}_0, \cdots, \mathbfit{G}_0 \right\}$ is a block diagonal matrix with $\mathbfit{G}_0$ on the diagonal, $\basicN\!=\!\CarN\CarBlocks$, and $\basicK\!=\!\CarK\CarBlocks$.

\subsection{Similarities and Differences}
From the previous two subsections, we see that both SC-LDPC codes and BMST codes can be derived from a small matrix by replacing the entries with properly-defined submatrices. We also see that the generator matrix $\mathbfit{G}_{\rm BMST}$ of BMST codes is similar in form to the parity-check matrix $\mathbfit{H}_{\rm SC}$ of SC-LDPC codes. SC-LDPC codes introduce memory by spatially coupling the parity-check matrices of the underlying LDPC-BCs, while BMST codes introduce memory by spatially coupling the generator matrices of the basic code. Thus, BMST codes can be viewed as a type of spatially coupled code. Similar to SC-LDPC codes, where increasing the lifting factor $M$ improves waterfall region performance, increasing the Cartesian product order $\CarBlocks$ of BMST codes also improves waterfall region performance. But the error floors, which are solely determined by the encoding memory $\Memory$~(see Section~\ref{SecIII_Bound}), cannot be lowered by increasing $M$.

\section{Performance Analysis of BMST Codes}\label{SecIII}
Throughout the paper, we consider binary phase-shift keying~(BPSK) modulation over the binary-input AWGNC. In this section, we first discuss the problem that prevents the use of conventional EXIT chart analysis for BMST codes, and then we provide a modified EXIT chart analysis to study the convergence behavior of BMST codes with window decoding.

\subsection{Genie-Aided Lower Bound on BER}\label{SecIII_Bound}
Let $p_b = f_{\rm BMST}(\gamma_b)$ be the BER performance function corresponding to a BMST code with encoding memory~(coupling width) $\Memory$ and coupling length $L$, where $p_b$ is the BER and $\gamma_b \triangleq E_b/N_0$ is in dB. Let $p_b = f_{\rm Basic}(\gamma_b)$ be the BER performance function of the basic code. By assuming a genie-aided decoder, we have~\cite{Ma13x}
\vspace{\vspaceh}
\begin{equation}\label{lowerbound}
    f_{\rm BMST}(\gamma_b) \!\geq\! f_{\rm Basic}\!\left(\! \gamma_b \!+\! 10\log_{10}\left(\Memory\!+\!1\right) \!-\!
    10\log_{10}\left(1 \!+\! \frac{\Memory}{L}\right) \!\right)\!,\!
\end{equation}
where the term $10\log_{10}\left(\Memory+1\right)$ depends on the encoding memory $\Memory$ and the term $10\log_{10}\left(1+\Memory/L\right)$ is due to the rate loss. In other words, a maximum coding gain over the basic code of $10\log_{10}(\Memory+1)$ dB in the low BER~(high signal-to-noise ratio~(SNR)) region is achieved for large $L$. Intuitively, this bound can be understood by assuming that a codeword in the basic code is transmitted $\Memory+1$ times without interference.

\subsection{A Modified EXIT Chart Analysis}
%by determining the minimum value of the SNR $E_b/N_0$ such that the mutual information~(MI) between the \emph{a posteriori} message at a variable node and an associated codeword bit~(referred to as the \emph{a posteriori MI} for short) goes to 1 as the number of iterations increases.

To describe density evolution, it is convenient to assume the all-zero codeword is transmitted and to represent the messages as log-likelihood ratios. The threshold of protograph-based LDPC codes can be obtained based on a protograph-based EXIT chart analysis~\cite{Liva07,Wei14_IT} by determining the minimum value of the SNR $E_b/N_0$ such that the MI between the \emph{a posteriori} message at a variable node and an associated codeword bit~(referred to as the \emph{a posteriori} MI for short) goes to 1 as the number of iterations increases, i.e., the BER at the variable nodes tends to zero as the number of iterations tends to infinity. However, as shown in~(\ref{lowerbound}), the high SNR performance of BMST codes with window decoding cannot be better than the corresponding genie-aided lower bound, which means that the \emph{a posteriori} MI of BMST codes does not tend to 1 as the number of iterations tends to infinity. Thus, the conventional EXIT chart analysis cannot be applied directly to BMST codes.

\begin{figure}[t]
  \centering
  \includegraphics[width=\FigOneWidth]{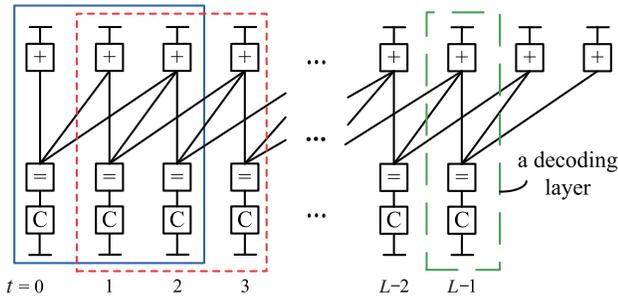}
  \caption{Example of a window decoder with decoding delay $d=2$ operating on the normal graph of a BMST code ensemble with $\Memory = 2$ at times $t=0$~(solid blue), and $t=1$~(dotted red). For each window position/time instant, the first decoding layer is called the target layer.}
  \label{BMST_graph}
\end{figure}

For convenience, the MI between the \emph{a priori} input and the corresponding codeword bit is referred to as the \emph{a priori MI}, the MI between the \emph{extrinsic} output and the corresponding codeword bit is referred to as the \emph{extrinsic} MI, and the MI between the channel observation and the corresponding codeword bit is referred to as the \emph{channel} MI. The analysis assumes that the interleavers $\mathbfit{\varPi}_i$~($0 \leq i \leq \Memory$) are arbitrarily large and random.

BMST code ensembles can be represented by a Forney-style factor graph, also known as a normal graph~\cite{Forney01}, where edges represent variables and vertices~(nodes) represent constraints. All edges connected to a node must satisfy the specific constraint of the node. A full-edge connects to two nodes, while a half-edge connects to only one node. A half-edge is also connected to a special symbol, called a ``dongle", that denotes coupling to other parts of the transmission system~(say, the channel or the information source)~\cite{Forney01}. There are three types of nodes in the normal graph of BMST codes.\footnote{For more details on the normal realization of BMST codes, we refer the reader to~\cite{Ma13x,Liang14c}.}
\begin{itemize}
  \item \textbf{Node} $\fbox{+}$: All edges~(variables) connected to node $\fbox{+}$ must sum to zero. The message updating rule at node $\fbox{+}$ is similar to that of the check node in the factor graph of a binary LDPC code. The only difference is that the messages on the half-edges are obtained from the channel observations.

  \item \textbf{Node} $\fbox{=}$: All edges~(variables) connected to node $\fbox{=}$ must take the same (binary) value. The message updating rule at node $\fbox{=}$ is the same as that of the variable node in the factor graph of a binary LDPC code.

  \item \textbf{Node} \fbox{C}: All edges~(variables) connected to node \fbox{C} must satisfy the constraint specified by the basic code. The message updating rule at node \fbox{C} can be derived accordingly, where the messages on the half-edges are associated with the information source.
\end{itemize}

The normal graph of a BMST code ensemble can be divided into \emph{layers}, where each layer typically consists of a node of type \fbox{C}, a node of type \fbox{=}, and a node of type \fbox{+}. Similar to SC-LDPC codes, an iterative sliding window decoding EXIT chart analysis algorithm with decoding delay $d$ working over a subgraph consisting of $d+1$ consecutive layers can be implemented to study the convergence behavior of BMST codes.\footnote{As with SC-LDPC codes, the decoding delay $d$ must be chosen several times as large as the encoding memory $m$ in order to achieve good performance.} The first layer in any window is called the {\em target layer}. An example of a window decoder with decoding delay $d=2$ operating on the normal graph of a BMST code ensemble with $\Memory = 2$ is shown in Fig.~\ref{BMST_graph}. In our modified EXIT chart analysis, the convergence check at node \fbox{C} is performed as follows.

\vspace{0.1cm}
\begin{algorithm}{Convergence Check at Node \fbox{C}}\label{alg:convergence}
\begin{itemize}
  \item Let $I_A$ denote the \emph{a priori} MI and $I_E$ denote the \emph{extrinsic} MI. Then the \emph{a posteriori} MI $I_{\rm AP}$ is given by
    \begin{equation}\label{I_APP}
        I_{\rm AP} = J(\sqrt{[J^{-1}(I_A)]^2+[J^{-1}(I_E)]^2}),
    \end{equation}
    where the $J(\cdot)$ and $J^{-1}(\cdot)$ functions are given in~\cite{Brink04}, $I_A$ is the \emph{a priori} MI, and $I_E$ is the \emph{extrinsic} MI. Suppose that the \emph{a posteriori} MI is Gaussian. As shown in Section~III-C of~\cite{Brink01}, an estimate of the BER $p_{est}$ is then given by
    \begin{equation}\label{I_APP_BER}
        p_{est} = \Qfun{J^{-1}(1-I_{\rm AP})/2},
    \end{equation}
    where
    \begin{equation}\label{eq:Qfunction}
        \Qfun{x} = \frac{1}{\sqrt{2\pi}}\int_{x}^{\infty}\exp\left\{-\frac{t^2}{2}\right\}dt.
    \end{equation}

  \item If the estimated BER $p_{est}$ is less than the preselected target BER, a local decoding success is declared; otherwise, a local decoding failure is declared.
\end{itemize}
\end{algorithm}

Given a channel parameter ${E_b}/{N_0}$, the channel MI is given by
\begin{equation}\label{J_ch}
    I_{ch}=J\left(\sqrt{8 R_{\rm BMST}\frac{E_b}{N_0}}~\right).
\end{equation}
The modified EXIT chart analysis algorithm of BMST codes can now be described as follows.

\vspace{0.1cm}
\begin{algorithm}{EXIT Chart Analysis of BMST Codes with Window Decoding}\label{alg:decoding}
\begin{itemize}
  \item {\bf{Initialization}:} All messages over those half-edges (connected to the channel) at nodes $\fbox{+}$ are initialized as $I_{ch}$ according to~(\ref{J_ch}), all messages over those half-edges (connected to the information source) at nodes \fbox{C} are initialized as 0, and all messages over the remaining (inter-connected) full-edges are initialized as 0. Set a maximum number of iterations $I_{\max} > 0$.

  \item {\bf{Sliding window decoding}:} For each window position, the $d+1$ decoding layers perform MI message processing/passing layer-by-layer according to the schedule
    \begin{equation*}
        \fbox{+} \rightarrow \fbox{$\Pi$} \rightarrow \fbox{=} \rightarrow
        \fbox{C} \rightarrow \fbox{=} \rightarrow \fbox{$\Pi$} \rightarrow \fbox{+}.
    \end{equation*}
    After a fixed number of iterations $I_{\max}$, make a convergence check at node \fbox{C} using Algorithm~1. If a local decoding failure is declared, then window decoding terminates; otherwise, a local decoding success is declared, the window position is shifted, and decoding continues. A complete decoding success for a specific channel parameter ${E_b}/{N_0}$ and target BER is declared if and only if all target layers declare decoding successes.
\end{itemize}
\end{algorithm}

Now we can denote the iterative decoding threshold ${({E_b}/{N_0})}^*$ of BMST code ensembles for a preselected target BER as the minimum value of the channel parameter ${E_b}/{N_0}$ which allows the decoder of Algorithm~1 to output a decoding success, in the limit of large code lengths~(i.e., $M\rightarrow \infty$).

\section{Numerical Results}
In the simulations to compute the window decoding thresholds of BMST codes, we set a maximum number of iterations $I_{\max}=1000$.

\ifCLASSOPTIONonecolumn
\begin{figure*}[htbp]
\fi
\ifCLASSOPTIONtwocolumn
\begin{figure}[htbp]
\fi
\centering
\subfigure[~]{
\label{Rate_sigma}
\includegraphics[clip, width=\FigTwoWidth]{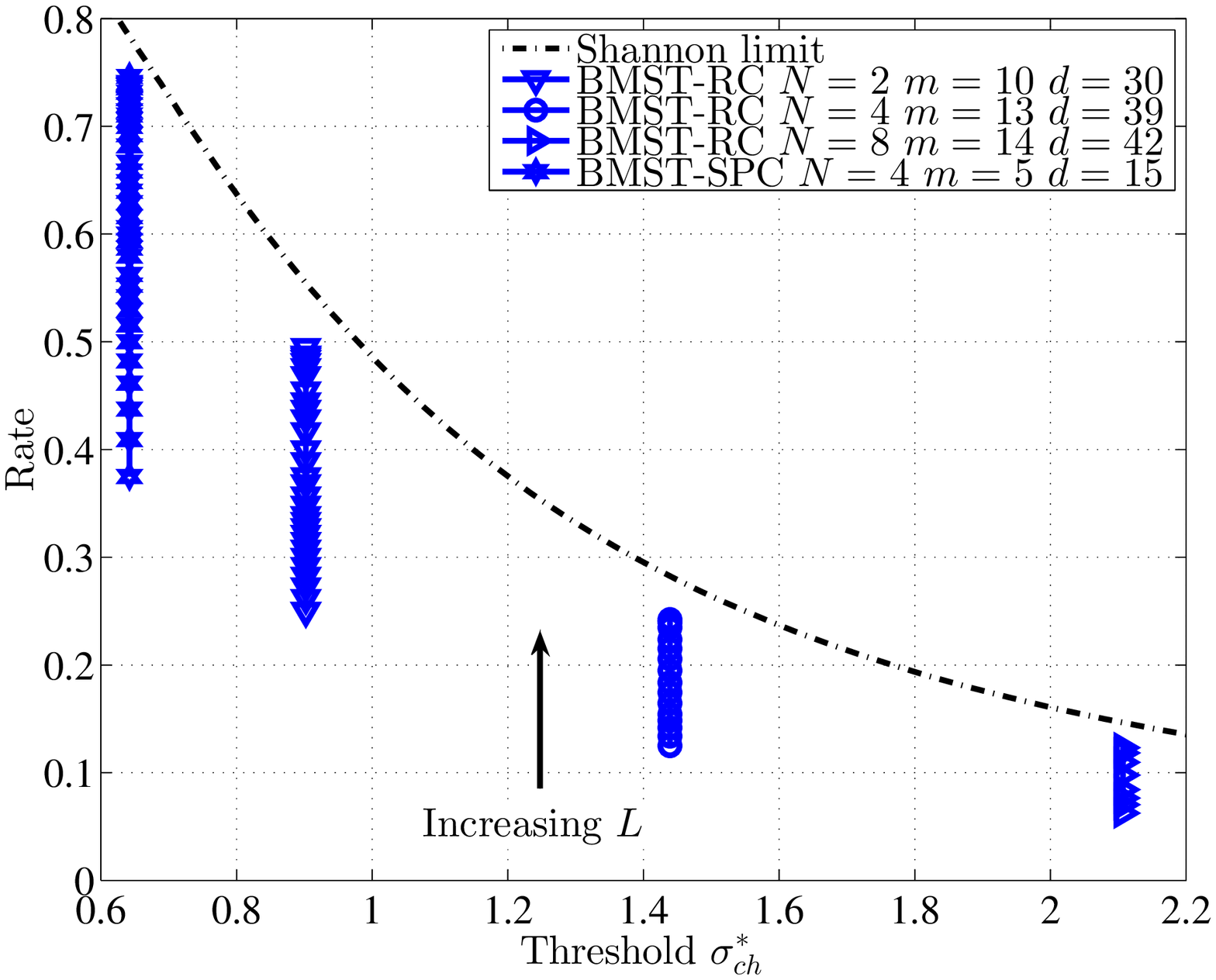}}
%\hspace{0in}
\subfigure[~]{
\label{Rate_EbN0}
\includegraphics[clip, width=\FigTwoWidth]{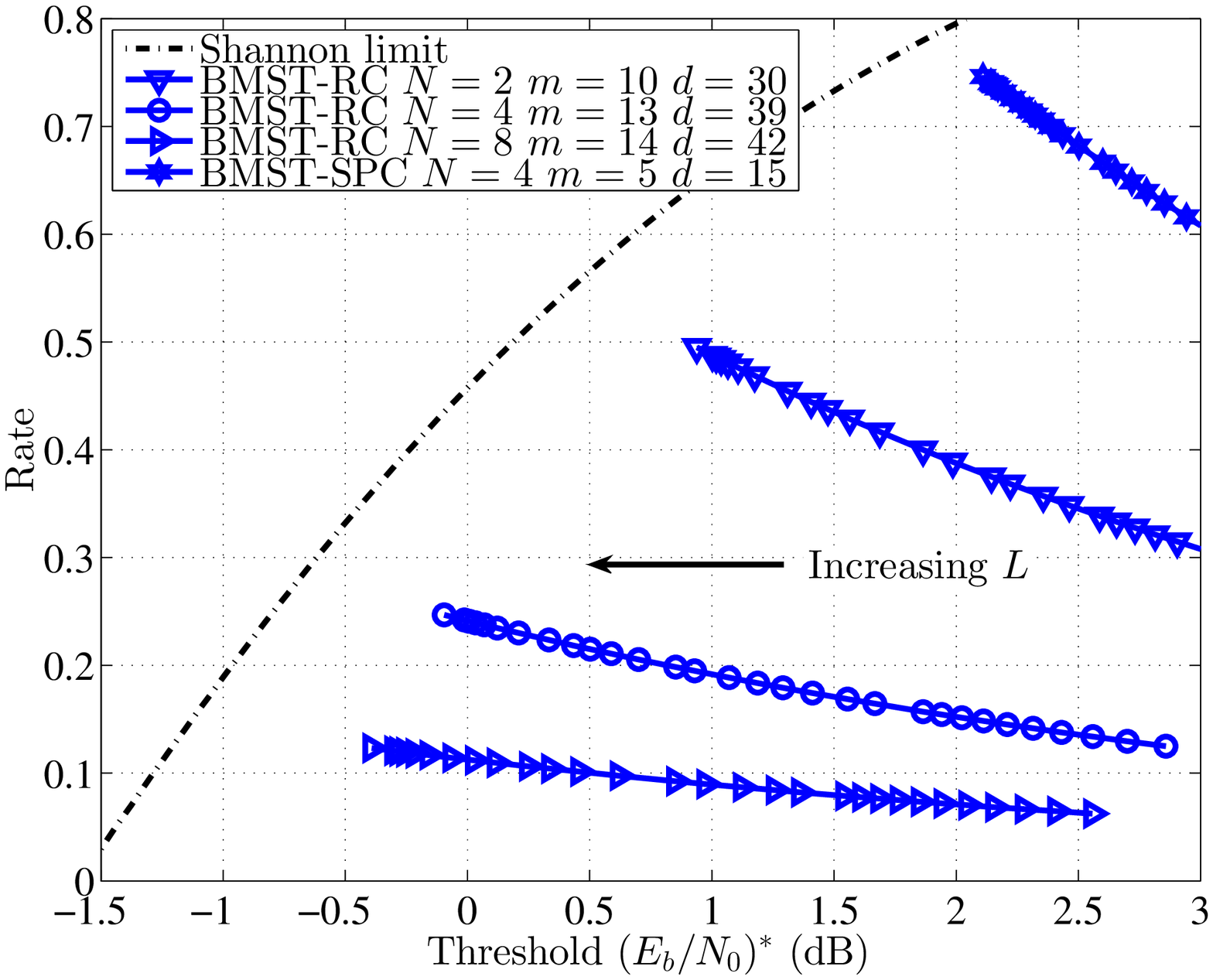}}
\caption{AWGNC BP thresholds in terms of (a) standard deviation $\sigma^{*}_{ch}$ and (b) SNR ${({E_b}/{N_0})}^*$ (dB) for several families of BMST code ensembles with increasing coupling length $L$, $m\leq L \leq 1000$, for a preselected target BER of $10^{-7}$.}
\label{Rate_sigma_EbN0}
\ifCLASSOPTIONonecolumn
\end{figure*}
\fi
\ifCLASSOPTIONtwocolumn
\end{figure}
\fi

\begin{example}
In Fig.~\ref{Rate_sigma_EbN0}, we display the thresholds of several families of BMST code ensembles with increasing coupling length $L$, $m\leq L \leq 1000$, for a preselected target BER of $10^{-7}$. The decoding delay\footnote{The threshold does not improve further beyond a decoding delay $d=3\Memory$.} is set to $d=3\Memory$. Fig.~\ref{Rate_sigma_EbN0}(a) plots the thresholds in terms of the standard deviation $\sigma^{*}_{ch}$ of the noise against the ensemble code rate $R_{\rm BMST}$. We observe that, as $L$ increases, the rate also increases while the threshold $\sigma^{*}_{ch}$ remains constant. The same thresholds are depicted in Fig.~\ref{Rate_sigma_EbN0}(b) in terms of the SNR ${({E_b}/{N_0})}^*$. Since ${E_b}/{N_0}$ takes into account the code rate, the thresholds ${({E_b}/{N_0})}^*$ improve monotonically with increasing $L$. However, in both plots, we can see that the gap to capacity decreases as $L$ increases.
\end{example}

\ifCLASSOPTIONonecolumn
\begin{figure*}[htbp]
\fi
\ifCLASSOPTIONtwocolumn
\begin{figure}[htbp]
\fi
\centering
\subfigure[BMST-RC codes]{
\label{RC21}
\includegraphics[clip, width=\FigThreeWidth]{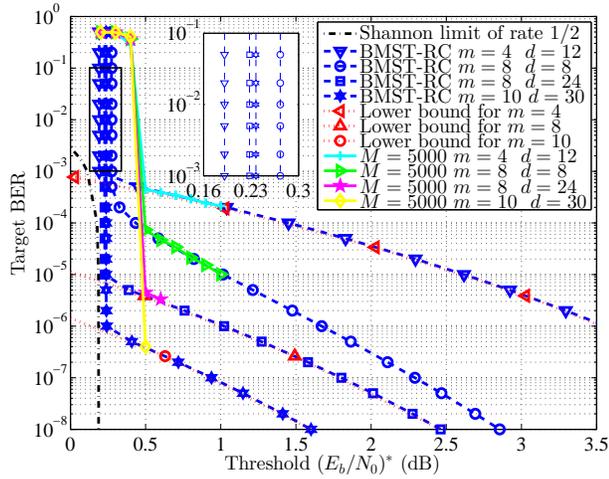}}
%\hspace{0in}
\subfigure[BMST-SPC codes]{
\label{SPC43}
\includegraphics[clip, width=\FigThreeWidth]{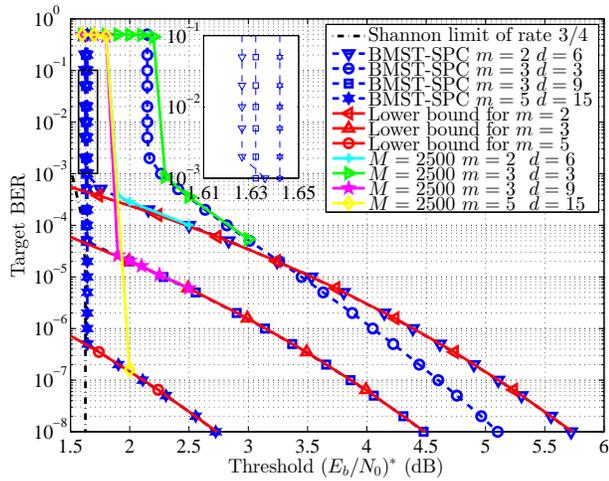}}
\caption{AWGNC BP thresholds in terms of ${({E_b}/{N_0})}^*$ (dB) for several families of BMST codes ensembles with different target BERs. The finite-length performance of BMST codes with RC $[2,1]^{5000}$ and SPC $[4,3]^{2500}$ as basic codes is also included. The coupling length $L=1000$.}
\label{Threshold_RC21_SPC43}
\ifCLASSOPTIONonecolumn
\end{figure*}
\fi
\ifCLASSOPTIONtwocolumn
\end{figure}
\fi

\begin{example}
For the coupling length $L=1000$, we calculated BP thresholds for several families of BMST code ensembles with different preselected target BERs. The calculated thresholds in terms of the SNR ${({E_b}/{N_0})}^*$ versus the preselected target BERs together with the lower bounds are shown in Fig.~\ref{Threshold_RC21_SPC43}, where we observe that
\begin{enumerate}
  \item For a fixed encoding memory $\Memory$, the thresholds remain constant at a value near capacity. Once a critical target BER is reached, however, the thresholds degrade rapidly as the target BER decreases further.
  \item For a high target BER~(roughly above $10^{-3}$), the threshold increases slightly as the encoding memory $\Memory$ increases, due to errors propagating to successive decoding windows.
  \item For a small decoding delay~(say $d=\Memory$), the thresholds do not achieve the lower bounds even in the high SNR region.
  \item For a larger decoding delay~(say $d=3\Memory$), the thresholds correspond to the lower bounds in the high SNR region, suggesting that the window decoding algorithm is near optimal for BMST codes.
  \item The error floor can be lowered by increasing the encoding memory $m$~(and hence the decoding delay $d$).
\end{enumerate}
In Fig.~\ref{Threshold_RC21_SPC43}, the window decoding performance of BMST codes with RC $[2,1]^{5000}$ and SPC $[4,3]^{2500}$ as basic codes is also plotted. By comparing the thresholds to the finite-length code performance, we conclude that the modified EXIT chart analysis for BMST codes is supported by the finite-length performance simulations. Note also that the gap between the simulated curves and the thresholds increases as the Cartesian product order $M$ of BMST codes decreases, as expected.
\end{example}

\section{Conclusions}\label{sec:Conclusion}
In this paper, we have proposed a modified EXIT chart analysis, that takes into account the relation between the MI and the BER, to calculate the window decoding thresholds of BMST codes. In this analysis, a BP algorithm is performed on the corresponding high-level normal graph of a BMST code ensemble. Using the modified EXIT chart analysis, we can predict the performance of BMST codes in the waterfall region of the BER curve. Finally, we showed that the EXIT chart analysis results are consistent with finite-length performance simulations.

%\section*{Acknowledgment}
%This work was partially supported by the $973$ Program (No. $2012$CB$316100$), the China NSF (No. 61172082), and the U.S. NSF (No. CCF-1161754). The authors are grateful to Mr. C. Liang for useful discussions.

\ifCLASSOPTIONcaptionsoff
  \newpage
\fi

%\bibliographystyle{IEEEtran}
%\bibliography{kechao}
%IEEE Trans.~Commun.
%IEEE Trans.~Inf.~Theory
%IEEE Int. Symp. on Inf. Theory
%Proc. Inf. Theory and App. Workshop
%\balance

\end{document}